\documentclass[10pt,a4paper,aps,prb,twocolumn,superscriptaddress,showpacs]{revtex4}

\usepackage[T1]{fontenc}
\usepackage[utf8x]{inputenc}

\usepackage[%
pdfauthor={Andreas J. Herrmann},%
pdftitle={Nonequilibrium Dynamical Cluster Approximation study of the Falicov-Kimball Model},%
]{hyperref}

\usepackage{dca}
\usepackage[capitalise]{cleveref}


\begin{document}

%

\title{
  Nonequilibrium Dynamical Cluster Approximation study \\
  of the Falicov-Kimball Model}


\author{Andreas J. Herrmann}
\email[]{andreas.herrmann@unifr.ch}
\affiliation{Department of Physics, University of Fribourg, 1700 Fribourg, Switzerland}
\author{Naoto Tsuji}
\affiliation{RIKEN Center for Emergent Matter Science (CEMS), Wako 351-0198, Japan}
\author{Martin Eckstein}
\affiliation{Max Planck Research Department for Structural Dynamics, University 
of Hamburg-CFEL, 22761 Hamburg, Germany}
\author{Philipp Werner}
\affiliation{Department of Physics, University of Fribourg, 1700 Fribourg, Switzerland}


\date{\today}

\begin{abstract}
    We use a nonequilibrium implementation of the dynamical cluster approximation (DCA) to study the effect of short-range correlations on the dynamics of the two-dimensional Falicov-Kimball model after an interaction quench. 
As in the case of single-site dynamical mean field theory, thermalization is absent in DCA simulations, and for quenches across the metal-insulator boundary,  nearest-neighbor charge correlations in the nonthermal steady state are found to be larger than in the thermal state with identical energy. We investigate to what extent it is possible to define an effective temperature of the trapped state after a quench.    
  Based on the ratio between the lesser and retarded Green's function 
  we conclude that a roughly thermal distribution is reached within the energy intervals corresponding to the momentum-patch dependent subbands of the spectral function. The effectively different chemical potentials of these distributions however lead to a very hot, or even negative, effective temperature in the energy intervals between these subbands. 
\end{abstract}


\maketitle

%


\section{Introduction}
The nonequilibrium dynamics of correlated fermionic lattice systems is of 
interest in connection with pump-probe experiments on solids, experiments on 
ultracold atoms in an optical potential and in the context of theoretical 
research on thermalization in many-body quantum systems. Over the last few 
years, the nonequilibrium extension of dynamical mean field theory (DMFT) 
\cite{freericks2006,aoki2014} has been developed into a powerful approach 
which allows to study the time evolution of high-dimensional lattice models. 
Applications of this method to the infinite-dimensional Hubbard model have 
produced interesting new insights, including, among others,
the transient trapping of the system in prethermalized 
states \cite{moeckel2008} after an interaction quench,\cite{eckstein2010} the 
existence of dynamical phase transitions,\cite{eckstein2009} the appearance 
of nonthermal critical points \cite{berges2004,tsuji2013,tsuji2013-a} and nonthermal order 
\cite{werner2012} in antiferromagnetic systems, as well as first order dynamical 
transitions in the Loschmidt echo.\cite{heyl2013,canovi2014}

While local time-dependent fluctuations can be accurately described within DMFT, the 
spatial degrees of freedom are treated at the mean-field level.
In low-dimensional systems, the effect of spatial fluctuations can be important for the dynamics, 
and to capture them, cluster extensions of nonequilibrium DMFT have been implemented.
The one- and two-dimensional Hubbard model has been studied within the 
dynamical cluster approximation (DCA) in Ref.~\onlinecite{tsuji2014}, using weak-coupling 
perturbation theory to solve the DMFT equations. In 
Ref.~\onlinecite{eckstein2014}, a four-site DCA calculation was used to 
simulate the effect of short-range antiferromagnetic correlations on the 
dynamics of a photo-doped Mott insulator. 
The relaxation rate of the photo-excited carriers was found to scale quadratically with the
nearest-neighbor spin correlations. 
In the latter study, the DCA 
equations were solved using a self-consistent strong-coupling perturbation 
theory (NCA).\cite{keiter1971,eckstein2010} 
At the moment, technical limitations prevent an extension of these methods to the intermediate coupling regime, where higher order versions of the strong-coupling expansion have to be used. Unbiased numerical methods, such as quantum Monte Carlo \cite{werner2009,werner2012} or 
DMRG \cite{wolf2014,balzer2015} are severely limited by an exponential scaling of the computational effort with the accessible time-range, and with cluster size. Hence, nonequilibrium DCA simulations of the Hubbard model are currently not only limited by the cluster size, which is essentially a memory issue, but most severely by the approximate methods used to solve the cluster impurity problem.  

In this study, we explore the effect of short-range correlations in 
the Falicov-Kimball (FK) model, \cite{falicov1969} which admits an exact 
solution within DMFT and DCA. While the dynamics of the FK model differs in 
many respects from that of the Hubbard model, due to the immobility of one 
spin-species, it exhibits a rich phase diagram in equilibrium, with metallic, 
Mott insulating and also long-range ordered phases. \cite{freericks2003} 
Hence, it is interesting to explore the effect of inter-site 
correlations on the relaxation properties of this model.

It is known from single-site nonequilibrium DMFT studies that the FK model does not thermalize after an interaction quench,\cite{eckstein2008} because on the one hand the distribution of immobile particles cannot adjust to the excited state of the system after the perturbation, and on the other hand the Hamiltonian of the mobile particles is quadratic for a given disorder configuration. In many situations involving the dynamics of quadratic Hamiltonians the relaxation results in non-thermal steady states which can be described by a generalized Gibbs ensemble (GGE).\cite{rigol2007-a} 
The latter takes into account constraints on the steady state in addition to energy and particle number conservation. A relevant question is therefore whether the trapped state obtained in DCA can be adequately described by a small number of effective parameters. In a first effort to address this question we study the energy distribution of the trapped states and investigate to what extent the distribution function can be characterized in terms of one or several temperatures and chemical potentials, and whether it is possible to extract a meaningful effective temperature which allows to explain the values of local and nonlocal observables. 

The rest of this paper is organized as follows. In Sec.~\ref{sec:model} we describe
the model and the implementation of the nonequilibrium DCA formalism. In Sec.~\ref{sec:eq}
we present equilibrium results for different cluster geometries, while Sec.~\ref{sec:noneq}
is devoted to the nonequilibrium results. Sec.~\ref{sec:conclusions} contains a brief
conclusion and outlook. 

\section{Model and Method}
\label{sec:model}
The Falicov-Kimball model \cite{falicov1969} was introduced
to 
describe semi-conductor metal transitions in SmB$_6$ and transition-metal 
oxides. It is similar to the Hubbard model \cite{hubbard1963} except that it 
distinguishes localized, and itinerant electrons. The Hamiltonian of the 
(spin-less) Falicov-Kimball model with nearest-neighbor hopping and local 
interactions reads
\begin{equation}
  \Ham = -t \sum_{\langle\rvec,\rvec'\rangle} \cdag_\rvec c_{\rvec'}
         + U \sum_\rvec \bigg(\cdag_\rvec c_\rvec - \frac{1}{2}\bigg)
                        \bigg(\fdag_\rvec f_\rvec - \frac{1}{2}\bigg),
  \label{eq:fkm_hamiltonian}
\end{equation}
where the $c$-electrons are itinerant, and the $f$-electrons are localized.  
Brandt {\it et.~al.}\cite{brandt1989,brandt1990,brandt1991} derived an exact 
solution of the Falicov-Kimball model in equilibrium in infinite dimensions 
using DMFT.\cite{metzner1989} Hettler {\it et.~al.} then introduced DCA 
\cite{hettler1998,hettler2000} as an extension to DMFT which takes non-local 
correlations into account and applied it to the Falicov-Kimball model in two 
dimensions. Also the nonequilibrium extension of DMFT was first applied to the 
Falicov-Kimball model. Freericks and coworkers discussed the damping of Bloch 
oscillations in the Falicov-Kimball model with static electric fields, 
\cite{freericks2006,freericks2008} while Eckstein and Kollar 
\cite{eckstein2008} studied its relaxation to a non-thermal steady state after 
an interaction-quench. Furthermore, Tsuji {\it et.~al.}\cite{tsuji2009} 
studied nonequilibrium steady-states in a driven Falicov-Kimball model using 
Floquet DMFT. In this work we use a nonequilibrium extension of the DCA 
formalism for the Falicov-Kimball model to compute the time-evolution of local 
and non-local observables after an interaction quench.

\begin{figure}[b]
  \centering
  \includegraphics{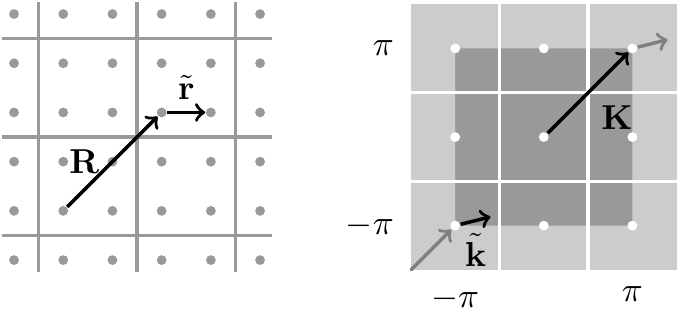}
  \caption{
  The sites in a cluster $\rcl$ form the unit-cell of a superlattice $\Rcl$ (left). 
  The first Brillouin zone of the original lattice is 
  split into patches of the size of the first Brillouin zone of the superlattice. The 
  corresponding reciprocal vectors $\kcl$ are centered around the reciprocal vectors $\Kcl$ 
  of the periodized cluster (right).
  }
  \label{fig:patch_constr}
\end{figure}

In a cluster extension of DMFT \cite{maier2005} one chooses a cluster of 
lattice-sites $\{\rcl\}$ such that the lattice maps to a super-lattice 
$\{\Rcl\}$ with the clusters as unit cells. In DCA, we then impose translation 
invariance under periodic boundary conditions on the cluster, which also leads 
to a renormalization of the hopping. The reciprocal vectors of the 
super-lattice $\{\kcl\}$ form patches around the reciprocal vectors of the 
cluster sites $\{\Kcl\}$, see \cref{fig:patch_constr}. Sites on the original 
lattice are decomposed as $\rvec = \Rcl + \rcl$, and points in the first 
Brillouin zone of the original lattice are decomposed as $\kvec = \Kcl + 
\kcl$.

\begin{figure}
  \centering
  \begin{tabular}{ccccc}
    $1{\times}1$ \\
    \includegraphics{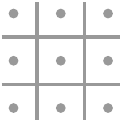}
    &
    \includegraphics{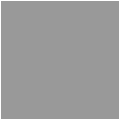}
    \\
    $1{\times}2$ & $a$ & & $b$ \\
    \includegraphics{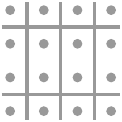}
    &
    \includegraphics{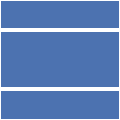}
    &
    &
    \includegraphics{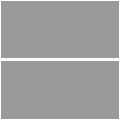}
    \\
    $1{\times}4$ & $a$ & & $b$ \\
    \includegraphics{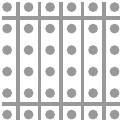}
    &
    \includegraphics{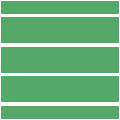}
    &
    &
    \includegraphics{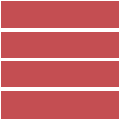}
    &
    \\
    $2{\times}2$ & $a$ & $b$ & $c$ & $d$ \\
    \includegraphics{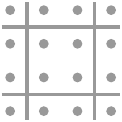}
    &
    \includegraphics{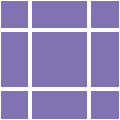}
    &
    \includegraphics{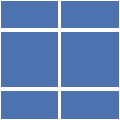}
    &
    \includegraphics{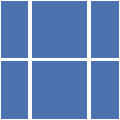}
    &
    \includegraphics{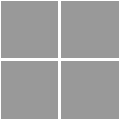}
    \\
    $2{\times}4$ & $a$ & $b$ & $c$ & $d$ \\
    \includegraphics{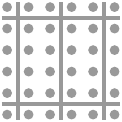}
    &
    \includegraphics{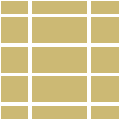}
    &
    \includegraphics{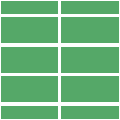}
    &
    \includegraphics{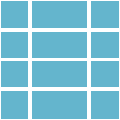}
    &
    \includegraphics{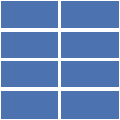}
    \\
    $2$ & $a$ & $b$ & $c$ & $d$ \\
    \includegraphics{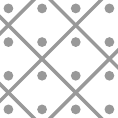}
    &
    \includegraphics{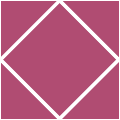}
    &
    \includegraphics{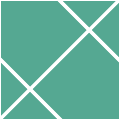}
    &
    \includegraphics{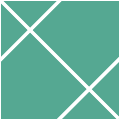}
    &
    \includegraphics{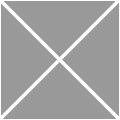}
    \\
    $8$ & $a$ & $b$ & $c$ & $d$ \\
    \includegraphics{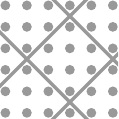}
    &
    \includegraphics{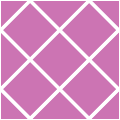}
    &
    \includegraphics{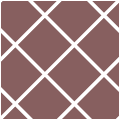}
    &
    \includegraphics{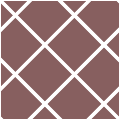}
    &
    \includegraphics{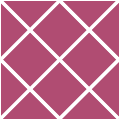}
    \\
  \end{tabular}
  \caption{Cluster geometries and patch-layouts considered in this work. The 
    left-most column depicts different cluster geometries in real-space. The 
    remaining columns depict some possible choices of patch 
    layouts in reciprocal space. Patch-layouts which are equivalent due to 
    symmetries in the dispersion $\epsilon_{\kvec}$ are grouped by color.
  }
  \label{fig:patches}
\end{figure}

The choice of reciprocal vectors $\{\Kcl\}$ is determined by the cluster 
shape.  However, we are free to choose the layout of the patches which 
associate the $\kvec$-vectors in the first Brillouin zone to the reciprocal 
vectors $\Kcl$. \Cref{fig:patches} depicts a number of cluster geometries and 
corresponding patch-layouts in reciprocal space. The left-most patch-layouts 
represent the canonical choice, where each $\kvec$-point is associated with 
the closest $\Kcl$ vector. Some of these patch-layouts are equivalent due to 
symmetries of the dispersion $\epsilon_{\kvec}$. For example, the layouts
$1{\times}1$, $1{\times}2b$, $2{\times}2d$, and $2d$ are equivalent.
Simulations on equivalent patch-layouts will yield identical results for 
observables on the whole system, even though the cluster-size and number of 
$f$-particle configurations might differ.

The Falicov-Kimball model (\ref{eq:fkm_hamiltonian}) maps to the following 
effective cluster impurity  Hamiltonian \cite{maier2005}
\begin{flalign}
   \Ham_\cl  - \mu \Num_\cl &= \Ham_0 + \Ham_f + \Ham_{\mathrm{int}} +
                      \Ham_{\mathrm{hyb}} + \Ham_{\mathrm{bath}},\phantom{\sum_{\Kcl,{\bf p}}} \\
   \Ham_0 &= \sum_\Kcl \epscl_\Kcl \cdag_\Kcl c_\Kcl -
             \mu \sum_\Kcl \cdag_\Kcl c_\Kcl, \\
   \Ham_f &= - \bigg(\frac{U}{2} + \mu\bigg) \sum_{\rcl} \fdag_{\rcl} f_{\rcl}, \\
   \Ham_{\mathrm{int}} &= \frac{U}{\Ncl}
                          \sum_{\Kcl,\Kcl',\rcl} \cdag_\Kcl c_{\Kcl'}
                          \bigg( \fdag_{\rcl}f_{\rcl} - \frac{1}{2} \bigg)
                          e^{-i(\Kcl-\Kcl')\rcl}, \\
   \Ham_{\mathrm{hyb}} &= \sum_{\Kcl,{\bf p}} \big( V_{\Kcl,{\bf p}}
                            \cdag_\Kcl a_{\Kcl,{\bf p}} + \mathrm{h.c.}
                          \big), \\
   \Ham_{\mathrm{bath}} &= \sum_{\Kcl,{\bf p}} \varepsilon_{\Kcl,{\bf p}}
                                      a^\dagger_{\Kcl,{\bf p}} a_{\Kcl,{\bf p}},
\end{flalign}
where $\mu$ is the chemical potential, $\Num_\cl$ the operator which counts the number of $c$ and $f$ particles, $c^{(\dagger)}$ and $f^{(\dagger)}$ are 
the (creation) annihilation operators for the mobile and localized electrons on the cluster, $a^{(\dagger)}$ are 
the bath (creation) annihilation operators, $V_{\Kcl,{\bf p}}$ are the hybridization parameters, and $\varepsilon_{\Kcl,{\bf p}}$ are the bath energy levels. We also introduce the dispersion of the lattice,  
\begin{equation}
\epsilon_{\kvec} = -\frac{1}{N} \sum_{\rvec\rvec'} e^{-i\kvec(\rvec-\rvec')}t_{\rvec\rvec'},
\end{equation}
and the patch averaged dispersion
\begin{equation}
  \epscl_{\Kcl} = \frac{\Ncl}{N} \sum_{\kcl} \epsilon_{\Kcl+\kcl},
\end{equation}
whose Fourier transform defines the hopping on the periodized cluster.

Furthermore, we introduce the $f$-particle configuration
\begin{equation}
  |\alpha\rangle = \Big(\prod_{\rcl} N^f_{\alpha \rcl} \fdag_{\rcl}\Big) |0\rangle, 
\end{equation}
where $N^f_{\alpha\rcl} \in \{0,1\}$ describes the $f$-particle occupation on 
cluster site $\rcl$. For a fixed $f$-particle configuration $\alpha$, we can 
trace out the bath states to obtain the cluster action
\begin{align}
    S^{\alpha}_\cl =
    &-\bigg(\frac{U}{2}+\mu\bigg) \sum_{\rcl}N^f_{\alpha\rcl} \nonumber\\
    &-\int_\cont d{z}\int_\cont d{z'} \sum_{\Kcl}
     \gamma^*_{\Kcl}(z) \Gweiss^{-1}_{\Kcl}(z,z') \gamma_{\Kcl}(z') \nonumber\\
    &+ \int_\cont d{z} \sum_{\Kcl,\Kcl'}
     \gamma^*_{\Kcl}(z) U_{\alpha\Kcl\Kcl'} \gamma_{\Kcl'}(z),
\end{align}
where $\gamma_\Kcl$ are the $c$-particle Grassman numbers,
\begin{equation}
  U_{\alpha\Kcl\Kcl'} =
    \frac{U}{\Ncl} \sum_{\rcl}
    \bigg( N^f_{\alpha\rcl} - \frac{1}{2} \bigg)
    e^{-i(\Kcl-\Kcl')\rcl}
\end{equation}
is the interaction matrix for configuration $\alpha$,
\begin{equation}
  \Gweiss^{-1}_{\Kcl}(z,z') =
  (i\deldel{z} - \epscl_\Kcl + \mu)\delta_\cont(z,z') - \Lambda_\Kcl(z,z')
\end{equation}
is the inverse excluded-cluster Green's function,\cite{maier2005} and
$\Lambda_\Kcl(z,z')$ is the hybridization function. The latter two are 
functions of two variables on the L-shaped contour $\cont$ which runs from 0 
to $t_{\max}$ and back on the real-time axis, and from 0 to $-i\beta$ on the 
imaginary-time axis.\cite{aoki2014}

The cluster Green's function is given by
\begin{align}
  G^\text{cl}_\Kcl(z,z') &= \sum_{\alpha}w_\alpha R_{\alpha\Kcl\Kcl}(z,z'), \label{eq:cluster}\\
  R_{\alpha\Kcl\Kcl'}(z,z') &= \frac{
  \int\text{D}[\gamma^*,\gamma][\T_\cont 
  e^{-S^{\alpha}_\cl}\gamma_\Kcl(z)\gamma^*_{\Kcl'}(z')]}{
  \int\text{D}[\gamma^*,\gamma][\T_\cont e^{-S^{\alpha}_\cl}]}, \\
  w_\alpha &= \frac{\int\text{D}[\gamma^*,\gamma][\T_\cont e^{-S^{\alpha}_\cl}]} {\sum_\alpha \int\text{D}[\gamma^*,\gamma][\T_\cont e^{-S^{\alpha}_\cl}]},
\end{align}
where the partial Green's function $R_{\alpha\Kcl\Kcl'}$ is the $c$-particle
Green's function for the fixed $f$-particle configuration $\alpha$, and 
$w_\alpha$ is the weight for this configuration. Evaluating the Gaussian 
Grassmann-integral yields the following contour Fredholm equation for the 
partial Green's functions in matrix notation:
\begin{align}
  [ \matGweiss^{-1} - \mat{U}_\alpha ]^{-1} = \mat{R_\alpha}, \quad
  [ \Id - \matGweiss\mat{U}_\alpha ] \mat{R}_\alpha &= 
  \matGweiss. 
  \label{eq:partial0}
\end{align}
Products indicate both matrix multiplication and contour convolution. The 
cluster Green's function and the excluded-cluster Green's function are 
diagonal in $\Kcl$, while the interaction matrix and the partial Green's function 
are not.

In DCA, we approximate the lattice self-energy by a piece-wise constant 
function in momentum space, whose values on the different momentum patches 
(Fig.~\ref{fig:patches}) are given by the cluster self-energy
$\Sigma_\Kcl=\Gweiss^{-1}_\Kcl-(G^\text{cl}_\Kcl)^{-1}$.
We use it to compute an approximate  lattice Green's function
\begin{equation}
G_{\Kcl+\kcl}(z,z') = [ i\deldel{z} + \mu - 
    \epsilon_{\Kcl+\kcl} - \Sigma_\Kcl]^{-1}(z,z'), \label{eq:lattice} 
\end{equation}
and the coarse-grained lattice Green's function
\begin{align}
  \bar{G}_\Kcl(z,z') = \frac{\Ncl}{N}\sum_{\kcl} G_{\Kcl+\kcl}(z,z'). 
\end{align}
The DCA self-consistency condition demands that $ \bar{\bf G}$ is identical to 
the cluster Green's function $\mat{G}^\text{cl}$. Thus, we can extract the new 
cluster-excluded Green's function by solving the Dyson equation
\begin{align}
  \matGweiss^{-1} - \mat{\bar G}^{-1} = \mat{\Sigma}, \label{eq:exclgf}
  \quad (\Id + \mat{\bar G}\mat{\Sigma})\matGweiss &= 
  \mat{\bar G}.
\end{align}

Obtaining the cluster self-energy from the cluster Green's function requires 
the introduction of helper functions (see \cref{sec:solvesigma}). In practice, 
we solve the equation
\begin{align}
(\Id + \mat{X}\matGweiss)\mat{\Sigma} &= \mat{X}, \label{eq:selfen} 
\end{align}
where $\mat{X}\matGweiss$ and $\mat{X}$ are given by
\begin{align}
  \mat{X}\matGweiss &= \sum_{\alpha} w_\alpha \mat{U}_\alpha \mat{R}_\alpha,  
  \\
  \mat{X} &= \sum_{\alpha} w_\alpha \mat{U}_\alpha\mat{R}_\alpha\mat{U}_\alpha.
\end{align}

In summary, the DCA solution for the Falicov-Kimball model consists of the 
following  steps: First we obtain a self-consistent solution for the initial 
equilibrium state.  Then we iteratively time-evolve from this equilibrium 
solution by solving the self-consistency loop at each time-step. The steps in 
the self-consistency loop are:
\begin{enumerate}
    \setcounter{enumi}{-1}
  \item Start from an initial guess for the self-energy $\Sigma_\Kcl(z,z')$ 
    (usually a zero-order extrapolation from the previous time-step).
  \item Solve the lattice problem (\ref{eq:lattice}) and compute the 
    coarse-grained lattice Green's function $\bar{G}_\Kcl(z,z')$.
  \item Obtain $\Gweiss_\Kcl(z,z')$ from Eq.~(\ref{eq:exclgf}).
  \item Solve the cluster problem and calculate $G^{\mathrm{cl}}_\Kcl(z,z')$ 
    from Eqs.~(\ref{eq:cluster}) and (\ref{eq:partial0}).\\
    (The configuration weights $w_\alpha$ only have to be calculated in
    the initial equilibrium state, since they are time-independent.)
  \item Obtain the self-energy $\Sigma_\Kcl(z,z')$ from \cref{eq:selfen}.
  \item Start over with step 1 until convergence.  Then start over with step 0 
    at the next time-step.
\end{enumerate}

\begin{figure}
  \includegraphics[width=\linewidth]{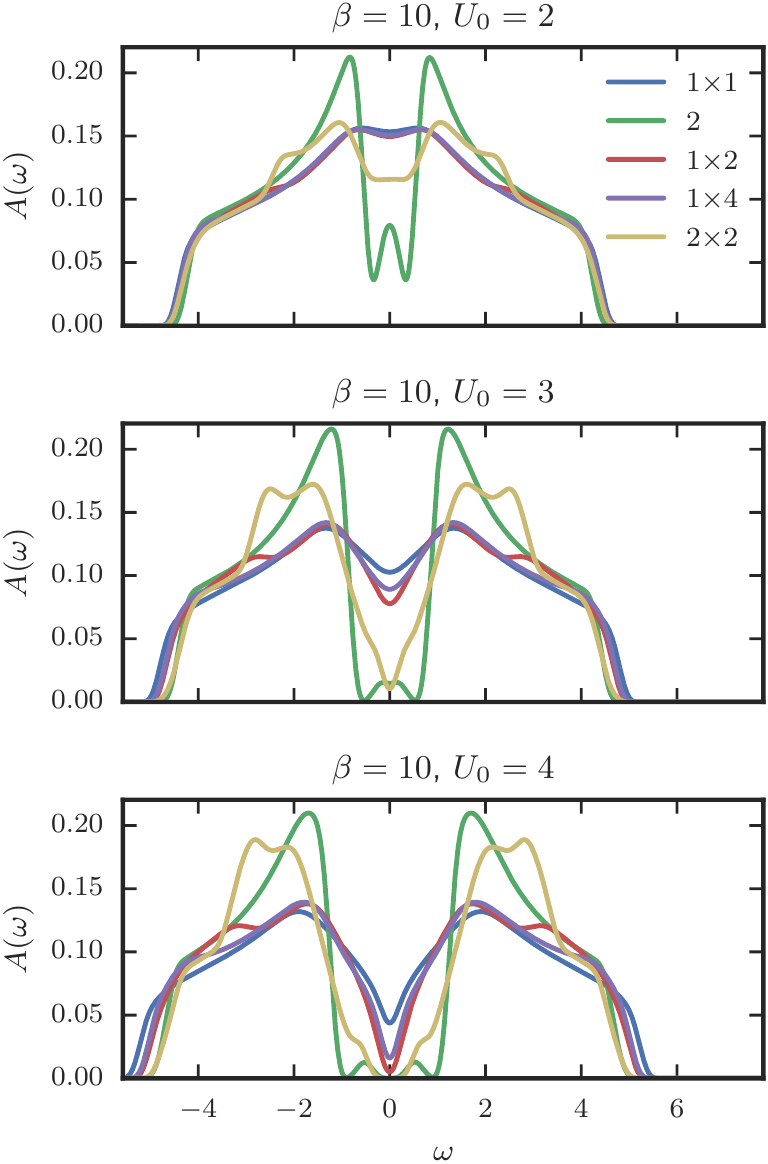}
  \caption{\label{fig:eq_spectrum}Comparison of the spectral function for 
  different cluster geometries and interaction parameters in equilibrium.
  In all cases the canonical patch-layout was used.
  }
\end{figure}

\begin{figure*}
  \includegraphics[width=\linewidth]{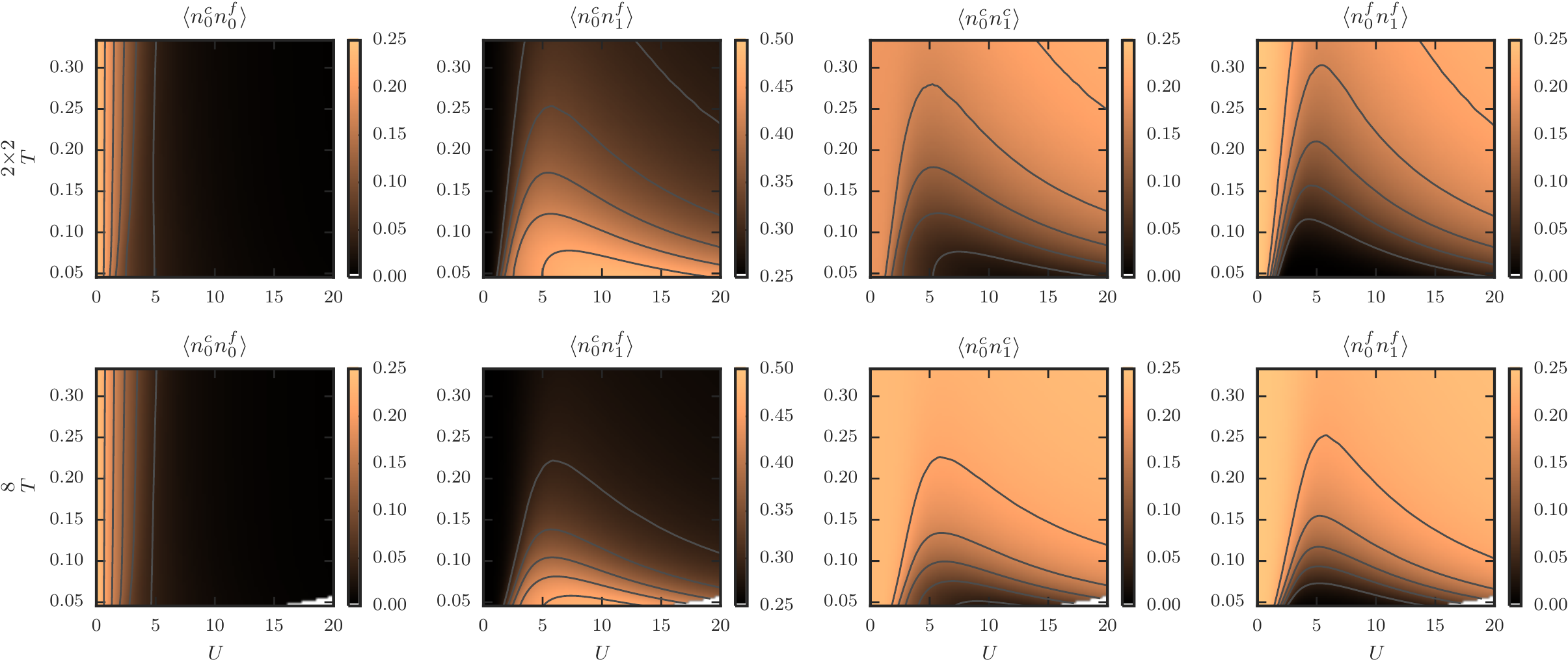}
  \caption{\label{fig:eq_phase_diag}Local and non-local cluster correlations 
    in equilibrium at different temperatures and interaction strengths. The 
    first column depicts the double occupation, the next three columns depict 
    nearest-neighbor density-density correlations between $c$- and $f$-, $c$- 
    and $c$-, and $f$- and $f$-particles, respectively. The upper row 
    corresponds to the $2{\times}2$ cluster, the lower row corresponds to the 
    $8$-site cluster. In both cases the canonical patch-layout was used.
  }
\end{figure*}

\begin{figure}
  \includegraphics[width=\linewidth]{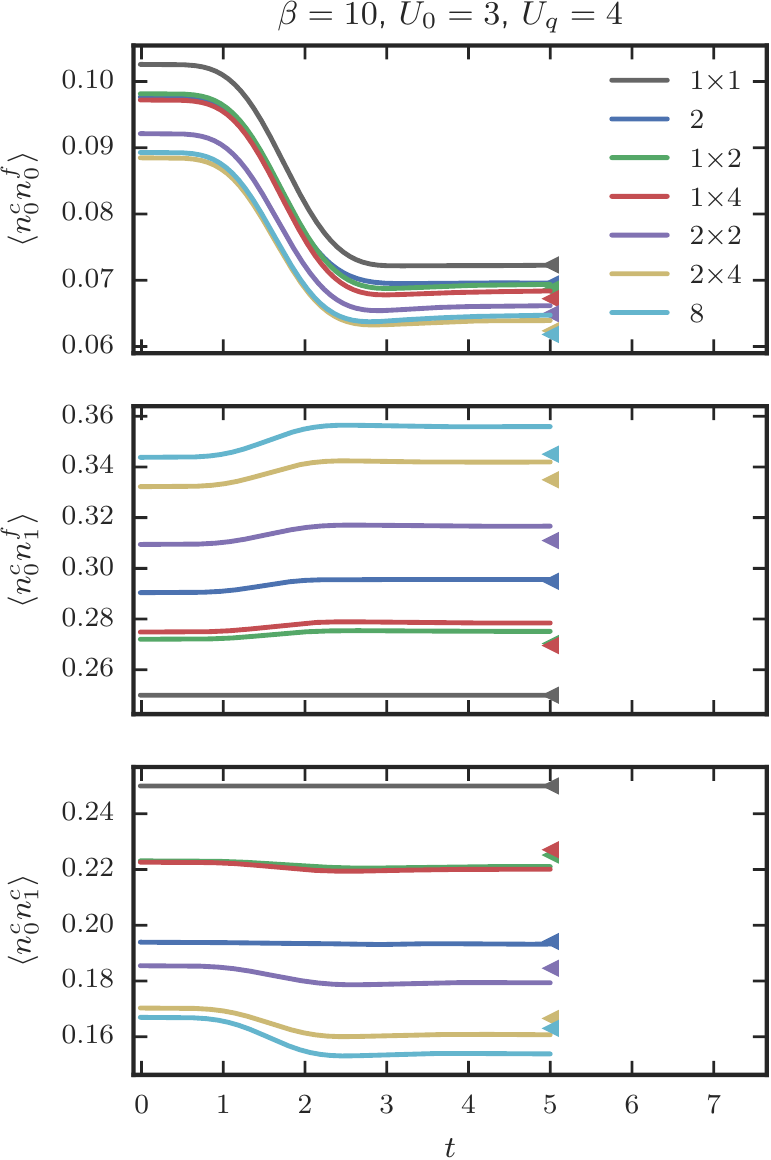}
  \caption{\label{fig:quench_correlations}Local and non-local correlations 
    between $c$-, and $f$-particles after an interaction ramp from $U_0=3$ to $U_q=4$, starting from an equilibrium state at $\beta=10$. 
    Results are shown for different cluster geometries, averaged over different 
    patch-layouts. The upper-most plot depicts the local double occupation.  
    The middle plot depicts nearest-neighbor density-density correlations 
    between $c$-, and $f$-particles. The lower-most plot depicts
    nearest-neighbor density-density correlations between $c$ particles. The
    triangles to the right indicate the expectation values in an equilibrium
    system with the same total energy.}
\end{figure}

\begin{figure}
  \includegraphics[width=.85\linewidth]{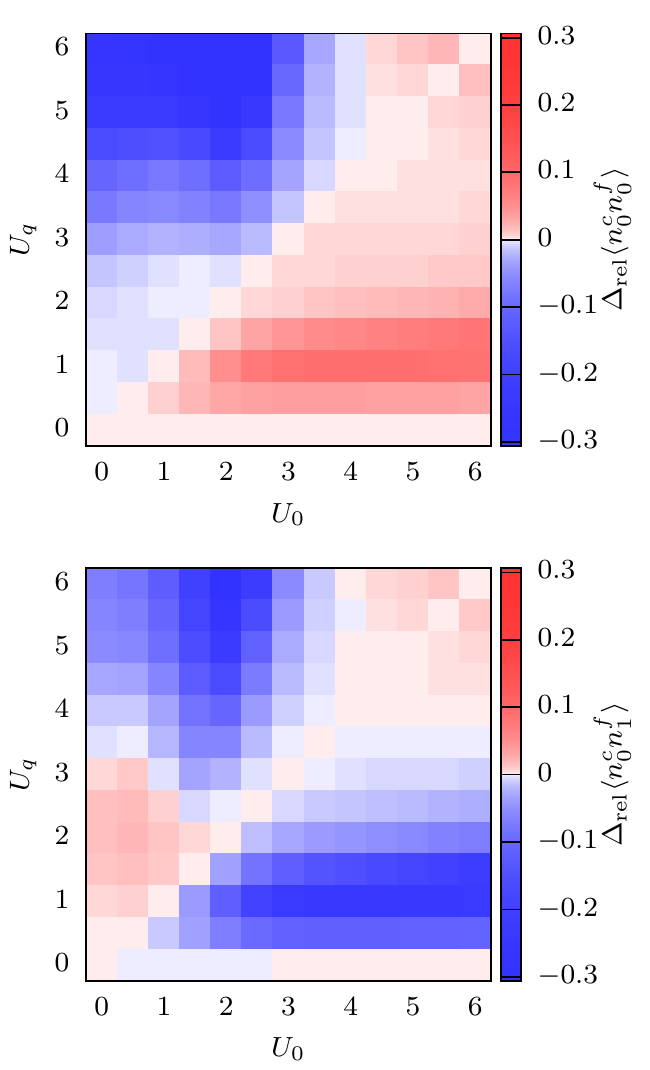}
  \caption{\label{fig:reldiff}
    Difference between the thermal expectation values and the steady-state observables (Eq.~(\ref{eq:delta_rel})) after an interaction ramp
    ($2{\times}2$ cluster, canonical patch-layout, initial $\beta=10$).
    The system is ramped from $U_0$ to $U_q$ according to Eq.~(\ref{eq:ramp}) and the double-occupancy
    and $c$-$f$ nearest-neighbor density-density correlation
    are measured. 
    The color scale in the double occupancy plot  is cropped to match the scale of the nearest-neighbor correlation plot.
  }
\end{figure}

\section{Equilibrium}
\label{sec:eq}

\subsection{Spectral Function}

In order to determine the spectral function (Fig.~\ref{fig:eq_spectrum}), we time-evolve the 
equilibrium system up to $t_\text{max}=30$, 
 Fourier-transform the retarded component of 
the Green's function
\begin{equation}
  G^R(\omega)
  = \int_{t_\mathrm{start}}^{t_\mathrm{max}}\de[t]\,
    G^R(t, t_{\mathrm{start}}) e^{i\omega (t - t_{\mathrm{start}})}
  ,
\end{equation}
where $t_{\mathrm{start}}=0$,
and use that
\begin{equation}
  \label{eq:four}
  A(\omega) = -\frac{1}{\pi}\Im G^R(\omega).
\end{equation}
We observe the opening of a gap in the range $3\lesssim U \lesssim 4$. The insulating nature is stronger when nonlocal correlations are included, which may be attributed to a charge ordering tendency (see next paragraph). 
However, the ``2'' cluster, and to a lesser extent the $2{\times}2$ cluster overestimate these charge order correlations, and hence the gap. 
Additional $K$-patches add features to the spectral function, some of which are artefacts of the piecewise-constant self-energy. 
Larger clusters than shown in Fig.~\ref{fig:eq_spectrum} would be needed for a converged solution.\footnote{Results for the 8-site cluster are not shown, because memory restrictions do not allow us to time propagate to $t=30$. Hence, we cannot reach the same spectral resolution as for the smaller clusters.}

\subsection{Local-, and Non-Local Observables}

The cluster Green's function contains non-local components and hence gives 
access to non-local observables. For example, we can calculate the nearest-neighbor 
density-density correlations between $c$-, and $f$-particles on the cluster as 
follows:
\begin{align}
  \langle n^c_{\rcl} n^f_{\rcl'} \rangle
  = \sum_\alpha w_\alpha \langle c^\dagger_{\rcl}c_{\rcl} \rangle_\alpha
                         N^f_{\alpha\,\rcl'}
  = \sum_\alpha w_\alpha \Im R^<_{\alpha\,\rcl\rcl} N^f_{\alpha\,\rcl'}.
\end{align}
Some cluster layouts break the symmetry between nearest-neighbor pairs along 
the horizontal or vertical axis. In order to mitigate this effect it is useful to average over
all nearest-neighbor pairs in the cluster, including those due to periodic 
boundary conditions. If there is no nearest neighbor along a given axis, as 
for example along the horizontal axis in the $1{\times}2$ cluster, then we 
apply the mean-field approximation $\langle n^c_{\rcl} n^f_{\rcl'} \rangle 
\approx \langle n^c_{\rcl} \rangle \langle n^f_{\rcl'} \rangle$. This way we 
obtain nearest-neighbor density-density correlations between $c$- and 
$f$-particles.

Nearest-neighbor density-density correlations between $c$-particles are obtained by applying Wick's theorem to the expectation value for each fixed $f$-particle configuration:
\begin{align}
  \langle n^c_{\rcl} n^c_{\rcl'} \rangle
  &= \sum_\alpha w_\alpha \langle
      c^\dagger_{\rcl}c_{\rcl} c^\dagger_{\rcl'}c_{\rcl'}
    \rangle_\alpha \nonumber\\
  &= \sum_\alpha w_\alpha \begin{multlined}[t] \big[
      \langle c^\dagger_{\rcl}c_{\rcl} \rangle_\alpha
      \langle c^\dagger_{\rcl'}c_{\rcl'} \rangle_\alpha \nonumber\\
    + \langle c^\dagger_{\rcl}c_{\rcl'} \rangle_\alpha
      \langle c_{\rcl}c^\dagger_{\rcl'} \rangle_\alpha
    \big] \end{multlined} \nonumber\\
  &= \sum_\alpha w_\alpha \begin{multlined}[t] \big[
      \Im R^<_{\alpha\,\rcl\rcl}\,\Im R^<_{\alpha\,\rcl'\rcl'} \\
    + \Im R^<_{\alpha\,\rcl'\rcl}(\delta_{\rcl\rcl'}-\Im R^<_{\alpha\,\rcl\rcl'})
    \big]. \end{multlined}
\end{align}
We apply the same averaging over nearest-neighbor pairs as described in the 
previous paragraph, including the mean-field approximation, if there are no 
nearest-neighbors along a certain axis.

Nearest-neighbor density-density correlations between $f$-particles are 
obtained in the same way. However, since there are no off-diagonal 
contributions to the occupation operator the expectation value simplifies to 
the following form,
\begin{equation}
  \langle n^f_{\rcl} n^f_{\rcl'} \rangle
  = \sum_\alpha w_\alpha N^f_{\alpha\,\rcl} N^f_{\alpha\,\rcl'}.
\end{equation}

It should be emphasised that due to the periodization in DCA, the 
nearest-neighbor 
cluster 
correlation functions are not identical to the corresponding lattice 
quantities. However, for large enough clusters, the
cluster correlations should provide a good estimate, so that 
it is meaningful to study the convergence with cluster size.

Local observables such as the double occupation can be obtained in the same 
manner:
\begin{equation}
  \langle n^c_{\rcl} n^f_{\rcl} \rangle
  = \sum_\alpha w_\alpha \Im R^<_{\alpha\,\rcl\rcl} N^f_{\alpha\,\rcl}.
\end{equation}
In the case of local quantities the DCA self-consistency condition guarantees 
that the cluster observables coincide with the lattice observables.

Equilibrium results of these correlation functions are depicted in 
Fig.~\ref{fig:eq_phase_diag} for the $2{\times}2$-, and the $8$-site cluster. The 
double occupation indicates a metal-insulator transition with weak temperature 
dependence, while the non-local correlations show a tendency towards charge 
order at low temperature, which is 
overestimated in the $2{\times}2$ case. 
We do
not explicitly break translation invariance. Therefore, we cannot observe an 
actual charge-order phase-transition. 
Nevertheless, our results are compatible
with the results by Hettler {\it et.~al.} \cite{hettler2000,freericks2003},
(for the symmetry broken phase)
except for the $2{\times}2$-site cluster, which shows the strongest 
suppression of charge-order in their study. Our correlation functions show the 
opposite effect, namely an enhancement of the charge order correlations in the canonical 
$2{\times}2$ geometry, which also explains the overestimation of the gap in Fig.~\ref{fig:eq_spectrum}. 

A more systematic trend with cluster size can be obtained by averaging over the different patch layouts shown in Fig.~\ref{fig:patches}.
After this averaging, the charge order correlations of the $2\times 2$ cluster become weaker than in the $8$-site cluster.

\section{Nonequilibrium}
\label{sec:noneq}

\subsection{Interaction Ramp}

In order to investigate the non-equilibrium dynamics of the system we start in an equilibrium state at finite temperature and change the interaction parameter according to the protocol  
\begin{align}
U(t) &= U_0 + (U_q - U_0)\; r\Big(\frac{t}{t_\mathrm{ramp}} - t_0\Big),
\label{eq:ramp}
\end{align}
with the ramp shape function
\begin{align}
  r(x) &= \begin{cases}
    0 & x < 0 \\
    \frac{1}{2} - \frac{3}{4} \cos(\pi x) + \frac{1}{4} \cos^3(\pi x)
      & 0 \le x < 1 . \\
    1 & 1 \le x
  \end{cases} 
  \label{eq:ramp2}
\end{align}
The ramp begins at $t_0$, and switches the interaction parameter from its initial value $U_0$ to the final value $U_q$ in a time $t_\mathrm{ramp}$. 
The smooth shape of the ramp function helps reduce the energy injected into the system.
Throughout this section we choose $t_0=0$ and $t_\mathrm{ramp}=3$.

The time-evolution of local and non-local correlation functions 
is shown in Fig.~\ref{fig:quench_correlations} for a ramp from $U_0=3$ to $U_q=4$. 
Here, we averaged the results over the different 
patch-layouts depicted in Fig.~\ref{fig:patches}. In the case of a 1-d Hubbard system, 
this type of averaging was found to improve the accuracy of the 
time-evolution.  \cite{tsuji2014}
Also for the present model and ramp set-up, it turns out that the averaging over patch-layouts results in a much more systematic 
trend with cluster size, even though an actual convergence cannot yet be observed with clusters up to 8 sites. 

The increase of $U$ moves the system further into the 
insulating regime, 
as confirmed by all three correlation functions,
and in particular by a reduction of the double occupancy.
The larger clusters exhibit both stronger initial nonlocal correlations and a stronger build-up of additional nonlocal correlations during the ramp. 
After the ramp the system relaxes to a non-thermal steady state. The small 
triangles depict the expectation values for an equilibrium system with the 
same total energy. If the non-equilibrium system were to thermalize, then the 
observables would converge to these results. Evidently the DCA simulations do 
not thermalize,
as expected for the Falikov-Kimball model, in which the distribution of 
$f$-particles
cannot react to the change in energy.  

While 
the reduction of the double-occupancy during the ramp is at least roughly consistent with the expected changes in a thermalizing system, 
the enhanced correlations in the nonlocal observables reflect a deviation from thermal equilibrium 
(apart from  the $1{\times}1$ cluster, where non-local correlations trivially factorize,  $\langle n^c_0n^f_1\rangle=\langle n^c\rangle \langle n^f\rangle$). 
To explain this effect and to
systematically investigate the deviation between the trapped state in the long-time limit and the corresponding equilibrium state with identical energy 
we have run a series of calculations for different initial interactions $U_0$ and final interactions $U_q$ for the $2{\times}2$ cluster at initial inverse temperature $\beta=10$.
As a local observable we consider the double occupation $\langle n^c_0n^f_0\rangle$ and as a non-local observable the $c$-$f$ nearest-neighbor correlation $\langle n^c_0 n^f_1\rangle$.
The expectation values of the trapped state are measured at $t=20$, and 
the relative difference to the thermal values is computed as
\begin{equation}
  \label{eq:delta_rel}
  \Delta_{\mathrm{rel}} O
  \equiv
  \frac{ O_{\mathrm{th}} - O }{ O_{\mathrm{th}} }.
\end{equation}

Figure~\ref{fig:reldiff} shows the measured deviations as intensity plots in the space of $U_0$ and $U_q$. 
Remembering that the critical interaction for the metal-insulator transition in the $2{\times}2$ cluster is about $U_c\approx 3$ (Fig.~\ref{fig:eq_spectrum}), we notice that for ramps within the metallic regime ($U_0,U_q<U_c$) 
or within the insulating regime ($U_0,U_q>U_c$)
the double-occpuation reaches a value close to the thermal one, while the thermal value substantially overestimates the trapped double-occupancy after up-ramps
across $U_c$. Similarly, the thermal value substantially underestimates the double-occupancy after down-ramps across $U_c$,  
except near $U_q=0$ where the correct double occupation of $0.25$ is reproduced.  

The result for $\langle n^c_0n^f_1\rangle$ is 
similar to that for the double-occupation, 
except that the sign of the relative change is opposite for the case of ramps into the metal regime.
For quenches across $U_c$ 
(independent of the quench direction) the short range charge-order is stronger in the nonthermal state than it would be after thermalization. This can be understood, because a higher temperature reduces the short-range correlations. Even in a coupling regime where a change of the interaction from $U_0$ to $U_q$  at fixed temperature would increase the charge correlations, the energy injected into the system can (if thermalized) more than compensate this and result in weaker correlations. One may furthermore wonder why nearest-neighbor charge correlations can increase at all after the quench (as in Fig.~\ref{fig:quench_correlations}), although the $f$-particles are static.  This can be explained because already in the initial state there is a short-range charge order, which implies nearest-neighbor {\em anti}-correlations between the $f$-particles. Since the $f$-particles are frozen, this short-range order remains after the quench. The nearest-neighbor $c$-$f$ correlations are hence likely to increase if an increase of $U$ leads to a reduction in the double occupancy, i.e. an increase in the density of $c$-particles at sites which are not occupied by $f$-particles. 

The 
sign change in $\Delta_\text{rel}\langle n^c_0n^f_0\rangle$ near $U_q=U_c$ ($U_c\approx 3$ is the critical value of the Mott transition) results in small deviations between thermal and trapped nearest-neighbor $c$-$f$ correlations after quenches to $U_q\approx U_c$. As we will see in the following section, this does however not mean that the energy distribution functions after such quenches are close to thermal distributions. 

\begin{figure*}[t]
  \includegraphics[width=.49\linewidth]{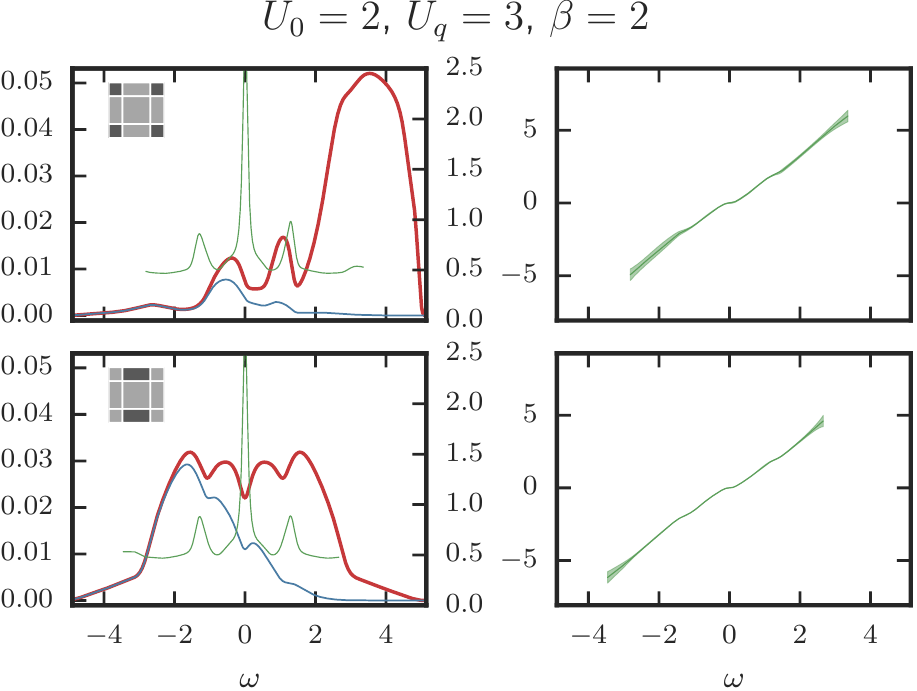}%
  \hfill
  \includegraphics[width=.49\linewidth]{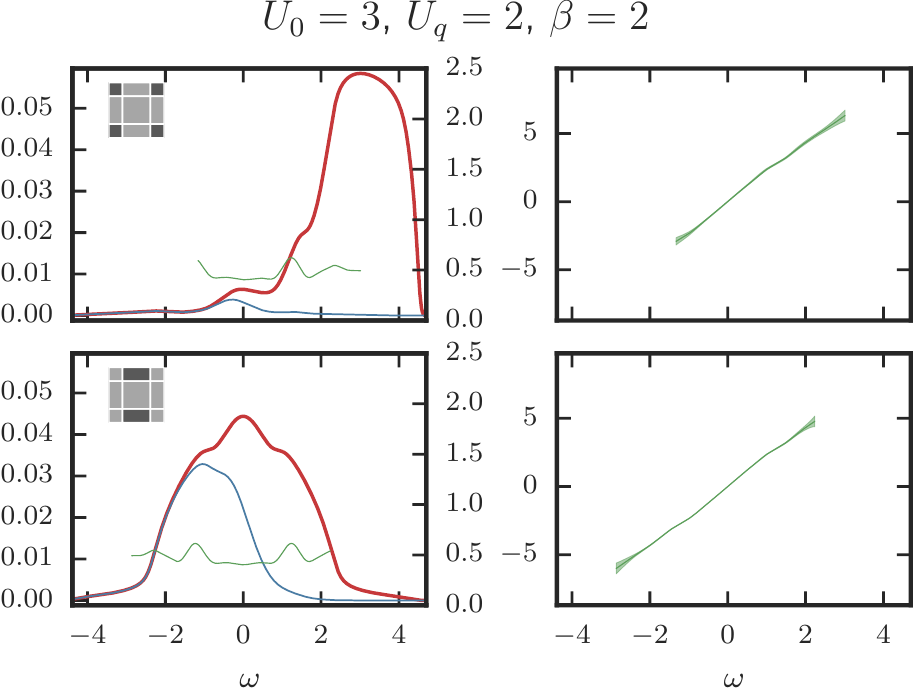} \\
  \includegraphics[width=.49\linewidth]{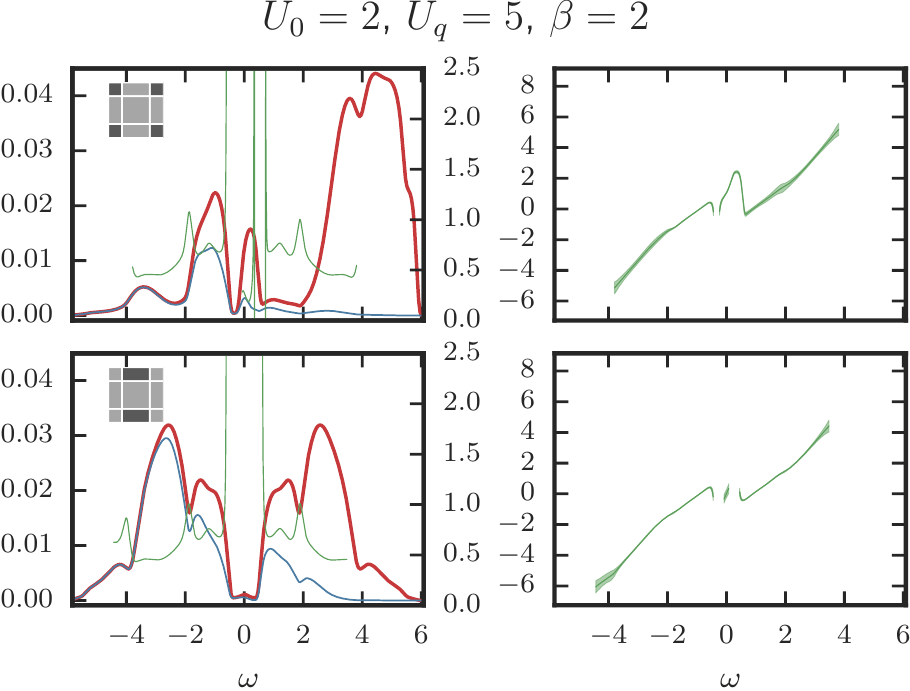}%
  \hfill
  \includegraphics[width=.49\linewidth]{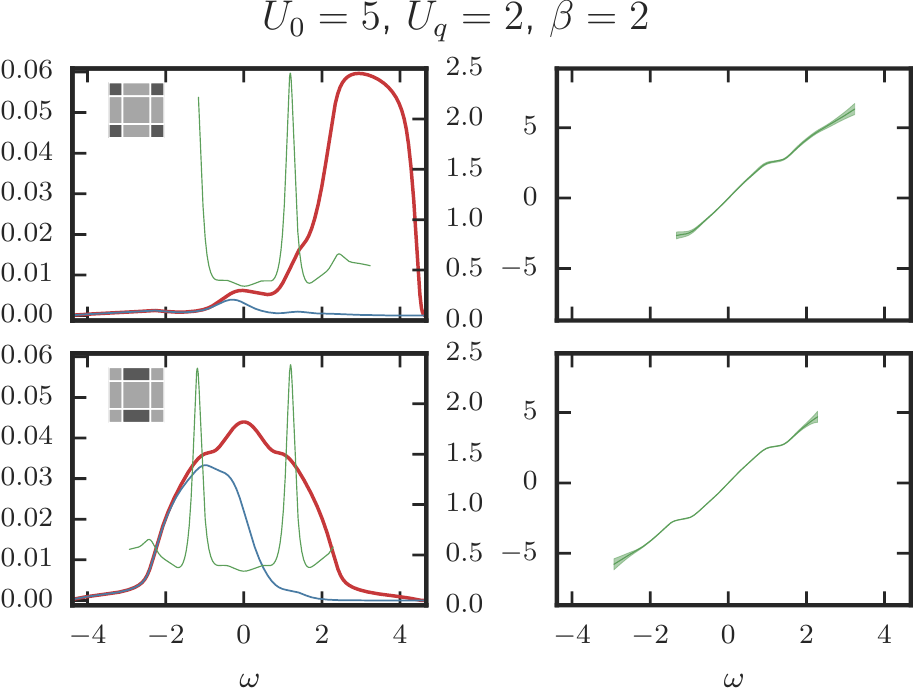}
  \caption{\label{fig:teff}
    Spectral- and occupation-functions for different $K$-patches after ramping the interaction parameter from $U_0$ to $U_q$ within the time interval $[0,3]$.
    The thick red line is the spectral function, and the thin blue line is the occupation function.
    The green line in the right panels depicts the function $h(\omega)$ whose slope would correspond to the inverse temperature $\beta$ in a thermalized system.
    The thin green line that is laid over the spectral function shows the reciprocal of that slope.
  }
\end{figure*}

\begin{table*}[ht]
\begin{ruledtabular}
\begin{tabular}{cclllllllll}

  cluster      & $U$               & $\beta$ & $\langle n^c_0 n^f_0 \rangle$ & $\langle n^c_0 n^f_1 \rangle$ & $\beta_{\mathrm{th}}$ & $\langle n^c_0 n^f_0 \rangle_{\mathrm{th}}$ & $\langle n^c_0 n^f_1 \rangle_{\mathrm{th}}$ & $\bar \beta_{\mathrm{eff}}$ & $\langle n^c_0 n^f_0 \rangle_{\mathrm{eff}}$ & $\langle n^c_0 n^f_1 \rangle_{\mathrm{eff}}$ \\
  \hline
  $2{\times}2$ & $2 \rightarrow 3$ & 2       & 0.121                         & 0.265                         & 1.56                  & 0.12                                        & 0.266                                       & 1.72                        & 0.117                                        & 0.268 \\
  $2{\times}2$ & $2 \rightarrow 5$ & 2       & 0.0992                        & 0.273                         & 0.905                 & 0.0856                                      & 0.261                                       & 1.45                        & 0.0633                                       & 0.273 \\
  $2{\times}2$ & $3 \rightarrow 2$ & 2       & 0.151                         & 0.268                         & 2.3                   & 0.152                                       & 0.264                                       & 2.09                        & 0.154                                        & 0.263 \\
  $2{\times}2$ & $5 \rightarrow 2$ & 2       & 0.149                         & 0.277                         & 2.36                  & 0.152                                       & 0.264                                       & 2.02                        & 0.154                                        & 0.263

\end{tabular}
\end{ruledtabular}
\caption{\label{tab:teff}
  Long-time expectation values and inverse temperatures 
  corresponding to Fig.~\ref{fig:teff}.
  The column $\beta$ indicates the inverse temperature before the ramp, $\beta_{\mathrm{th}}$ is the inverse temperature at which an equilibrium system with otherwise identical parameters has the same total energy as the system after the ramp, while 
  $\bar\beta_{\mathrm{eff}}\equiv 1/\bar T_\text{eff}$ is obtained from the mode (most common value) of the inverse temperatures extracted
  from the slope of $h(\omega)$. 
}
\end{table*}

\subsection{Effective temperatures}

In the absence of thermalization, an interesting issue is whether or not the state of the system can be characterized by a small number of parameters, such as effective temperatures or effective chemical potentials. In fact, since the Falicov-Kimball lattice model has a large number of conserved quantities, an exact description in terms of a generalized Gibbs ensembles (GGE) is possible (see Appendix \ref{sec:gge_appendix}). However, a GGE description with an extensive number of parameters is not very useful, and it is also not clear how this construction can be adapted to the DCA case. 

In DCA, the $f$-particle configurations are conserved by the time-evolution and one possible goal could be to devise a GGE-like description of the trapped state 
which is based 
on effective temperatures and chemical potentials that depend on the $f$ configuration. 
With this motivation in mind, we will investigate in the following sections to what extent the notion of an effective ($c$-electron) temperature is useful to characterize the trapped states observed in DCA simulations.

To address this issue, we consider the quantity 
\begin{equation}
h(\omega)=\log[-2 \Im G^R(\omega)/\Im G^<(\omega)-1],
\label{teff}
\end{equation}
which turns out to be independent of the real-space components or cluster momenta.
In thermal equilibrium, one has  $G^<(\omega) = 2\pi i A(\omega) f(\omega)$ due to the fluctuation-dissipation theorem,  so that $h(\omega)$
will linearly increase with a slope given by $\beta=1/T$. 
In the nonequilibrium case the slope of $h(\omega)$ yields a possible definition of an effective inverse temperature $\beta_\text{eff}\equiv 1/T_\text{eff}$.
In Eq.~(\ref{teff}), we do not show a time argument because we assume that the spectral functions are computed in the nonthermal steady state reached after the quench.
(For the results shown in Fig.~\ref{fig:teff}, we have propagated the solution up to $t=80$ and computed the spectral functions by Fourier transformation over the time-interval $[40,80]$.)

In the following, we focus on the 2$\times$2 cluster and choose a relatively high initial temperature $\beta=2$. Figure~\ref{fig:teff} 
plots the spectral function, occupation function 
and the quantity (\ref{teff}) for quenches from $U_0=2$ to $U_q=3$, $U_0=3$ to $U_q=2$,  $U_0=2$ to $U_q=5$ and $U_0=5$ to $U_q=2$. The
two panels correspond to the $k=(\pi,\pi)$ and $k=(0,\pi)$ components. 
Error bars on $h$ were estimated by error-propagation from the error $\sigma_A$ on the spectral functions,
which is mainly due to to the finite time-interval of the Fourier transform.
(The spectral function should integrate to 1 and we used the deviation in that integral to estimate $\sigma_A$.)
From the slope of $h(\omega)$ we extract the ``energy dependent effective temperature'' 
$[\de[h(\omega)]/\de[\omega]]^{-1} = T_\text{eff}(\omega)$, which is overlaid on the spectra in the left panels (green curves, right scale).  
Within the accuracy of our calculation, the effective temperatures are the same for all $k$ patches. 
After the quench from $U=3\rightarrow 2$, $h(\omega)$ exhibits an approximately linear $\omega$-dependence, roughly consistent with a thermal distribution, although there are flat regions near $\omega=\pm 1.5$ (resulting in a noticeable increase of the inverse slope $[\de[h(\omega)]/\de[\omega]]^{-1}$). After the $U=5\rightarrow 2$ ramp we observe pronounced spikes in $[\de[h(\omega)]/\de[\omega]]^{-1}$ at the same energies. Similarly, after the quenches from $U=2$ to larger interactions, the effective temperature profile shows large variations as a function of $\omega$. In particular, we note that even after the $U=2\rightarrow 3$ ramp, for which the local and nonlocal observables shown in Fig.~\ref{fig:reldiff} are close to their thermal counterparts, $T_\text{eff}(\omega)$ exhibits a large $\omega$-dependence, and therefore the steady state cannot be described by a single effective temperature.    

In general, one observes that the $h(\omega)$ curves feature offsets between different almost-linear intervals.
In each frequency interval with a linear slope of $h(\omega)$ the distribution function could be parametrized by a Fermi function with some effective chemical potential. 
Connecting two regions with different chemical potential offsets gives rise to plateaus in $h(\omega)$,
which in turn cause spikes in the ``temperature profile''.

\begin{figure*}[!t]
  \includegraphics[width=.50\linewidth]{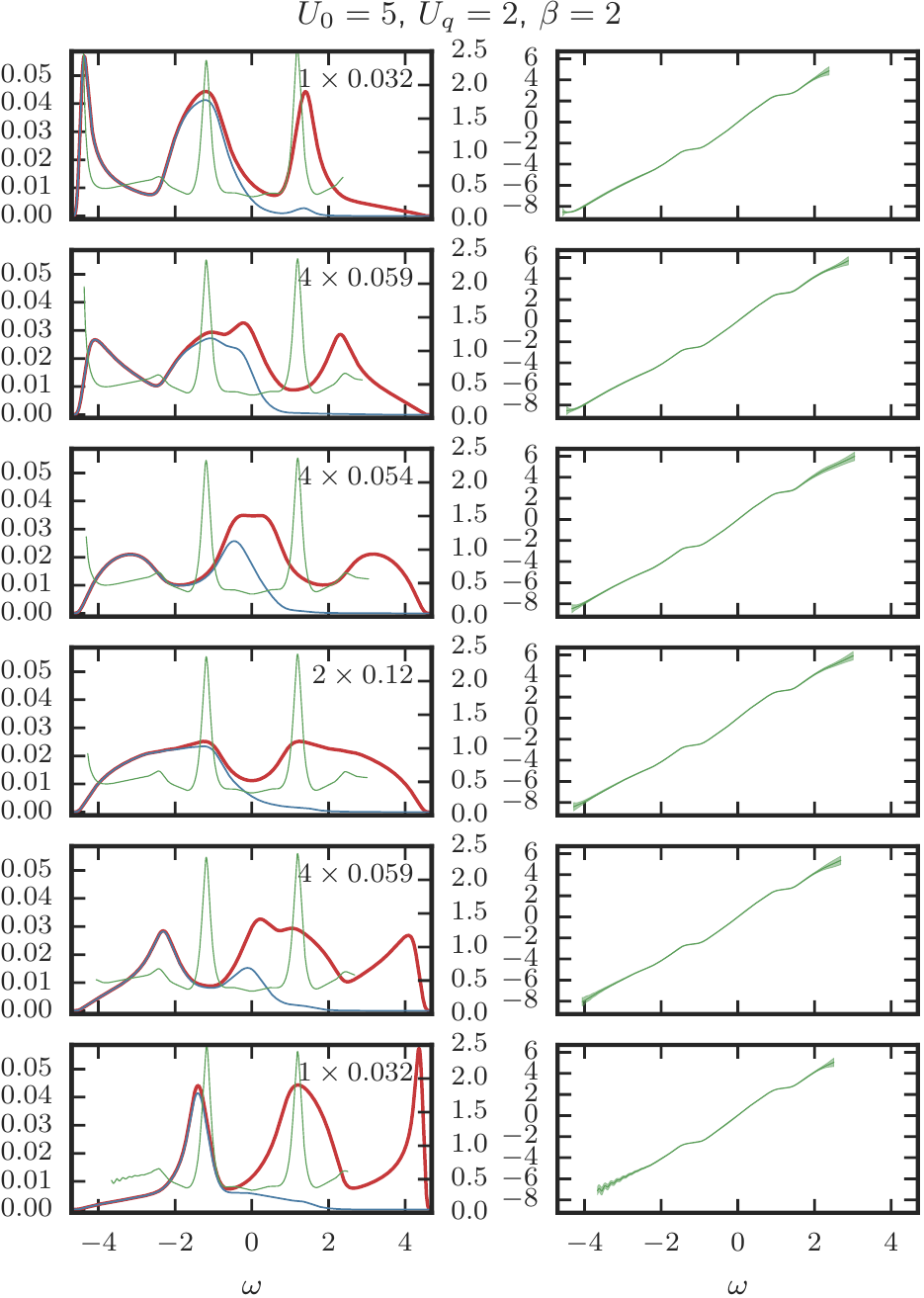}
  \includegraphics[width=.23\linewidth]{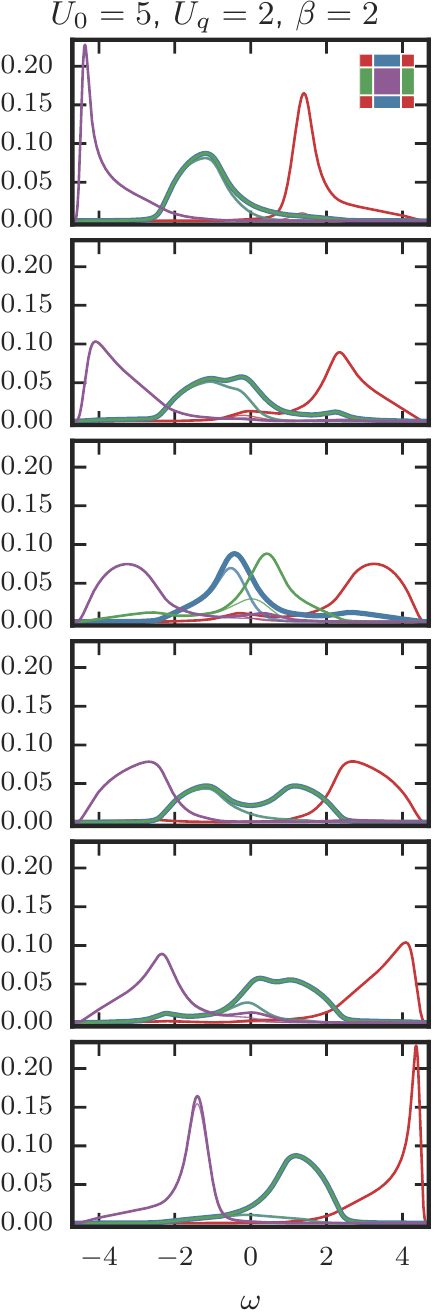}
  \includegraphics[width=.23\linewidth]{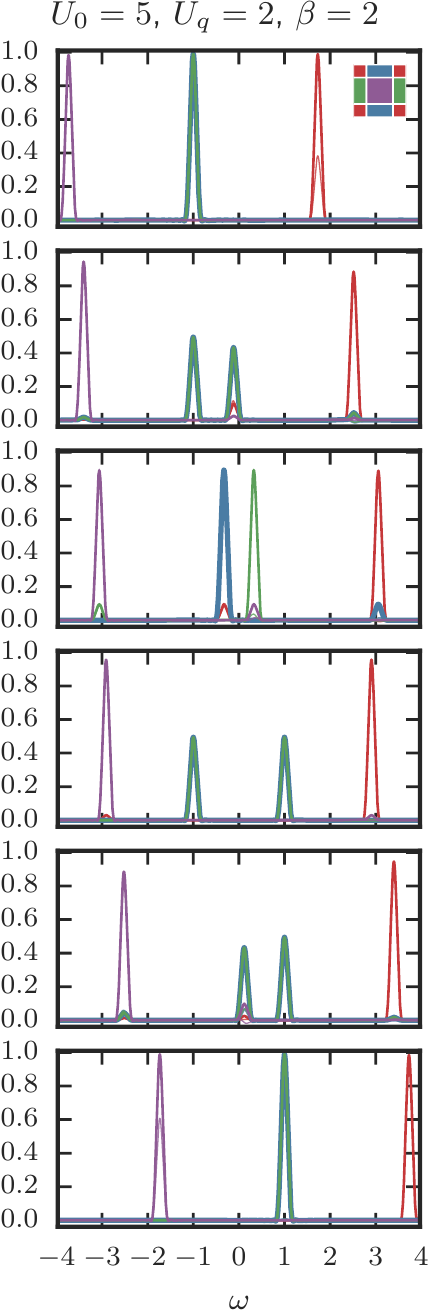}
  \caption{\label{fig:teff_conf}%
    First column:
    Spectral- and occupation-functions for different $f$-particle configurations after ramping the interaction parameter from $U_0$ to $U_q$ within the time interval $[0,3]$.
    The thick red line is the spectral function, and the thin blue line is the occupation function.
    Second column:
    Function $h(\omega)$, 
    whose slope would correspond to the inverse temperature $\beta$ in a thermalized system.
    The thin green line that is laid over the spectral function in the left panels shows the reciprocal of that slope,
    i.e.  $T_\text{eff}(\omega)$ (right axis).  
    The following $f$-particle configurations are depicted from top to bottom:
    Unoccupied, one $f$-particle, two particles along an edge, two particles along a diagonal, three $f$-particles, fully occupied.
    The numbers in the upper right corner of each panel in the first column denote the multiplicity of that configuration due to symmetry and the corresponding configuration weight $w_\alpha$.
    Third column:
    Spectral- and occupation-functions for different $f$-particle configurations and momentum patches.
    The spectra in the first column are the normalized sum of these.
    Fourth column:
    Energy spectrum for different $f$-particle configurations and momentum patches in an isolated cluster at $U=2$.
    The $(0,\pi)$, and $(\pi,0)$ spectra (green and blue curves) overlap in 
    all but the third row.
  }
\end{figure*}

By taking the mode (most common value) of $\beta_\mathrm{eff}(\omega)$
over the energy range in which this quantity can be accurately determined,
we obtain $\bar \beta_\mathrm{eff}\equiv 1/\bar T_\mathrm{eff}$, which may be regarded as a possible definition of the effective temperature of the trapped state.
The modal value is not sensitive to the pronounced spikes in $[\de[h(\omega)]/\de[\omega]]^{-1}$.
In Tab.~\ref{tab:teff} we compare the double occupancies and $\langle n^c_0 n^f_1 \rangle$ expectation values at  $\bar \beta_\mathrm{eff}$ to the trapped values,
and to the thermal values which the system would reach if it could thermalize at the given energy.
It is evident that  $\bar \beta_\mathrm{eff}$ does not provide a particularly accurate description of the observables in the trapped state;
the predictions based on $\bar \beta_\mathrm{eff}$ are generally worse than those based on the effective thermal temperature $T_\mathrm{th}=1/\beta_\mathrm{th}$.
This poor result is probably related to the fact that our effective thermal description ignores the fact that 
$h(\omega)$ is only piecewise (approximately) linear, i.e. different energy intervals have different effective chemical potentials. 

To get more insights into the $\omega$-dependence of $T_\text{eff}$ we consider the ramp from $U=5\rightarrow 2$,
which yields large spikes at the energies $\omega \approx \pm 1.2$ (Fig.~\ref{fig:teff}),
and calculate the contributions to the spectral function from the different $f$-particle configurations. 
Fig.~\ref{fig:teff_conf} shows the results for the following configurations (from top to bottom): 
No $f$-particles, one $f$-particle, two particles along an edge, two particles along a diagonal, three $f$-particles, fully occupied.
 
We observe that the frequency-dependent effective temperature is identical for all configurations,
and hence identical to the (momentum-independent) effective temperature in Fig.~(\ref{fig:teff}).
One can indeed show rigorously that if the distribution function of the cluster Green's function is independent of  momentum, then also the Green's function $\mat{R}_\alpha$ for each individual $f$-particle configuration must have the same distribution function. 
Let us assume that the system has relaxed to a steady-state in which the 
contour objects $\matGweiss$, and $\mat{R}_\alpha$ can be characterized 
by $\omega$-dependent retarded, advanced, and Keldysh components.
We can then use the parametrization
\begin{align}
  \label{eq:keldyshdist}
  \matGweiss^K &= \matGweiss^R \mat{F}_\Gweiss  + \mat{F}_\Gweiss \matGweiss^A, \\
  \mat{R}_\alpha^K &= \mat{R}_\alpha^R \mat{F}_\alpha  + \mat{F}_\alpha \mat{R}_\alpha^A,\label{RalphaK}
\end{align}
where $\mat{F}$ is the non-thermal distribution function.
Inserting the ansatz (\ref{eq:keldyshdist}) into Eq.~(\ref{eq:partial0}), we obtain the Keldysh component of the partial Green's function,
\begin{align}
  \mat{R}_\alpha^K
  &= \mat{R}_\alpha^R {\matGweiss^R}^{-1} \matGweiss^K {\matGweiss^A}^{-1} \mat{R}_\alpha^A \\
  &= \mat{R}_\alpha^R \mat{F}_\Gweiss - \mat{F}_\Gweiss \mat{R}_\alpha^A
  + \mat{R}_\alpha^R [\mat{F}_\Gweiss, \mat{U}_\alpha] \mat{R}_\alpha^A,
\end{align}
where  $[\cdot,\cdot]$ is the commutator.
If the distribution function $\mat{F}_\Gweiss$ is momentum-independent, then $\mat{F}_\Gweiss$ 
is proportional to the identity matrix and commutes with the interaction 
matrix $\mat{U}_\alpha$.
It then follows from comparison to Eq.~(\ref{RalphaK}) that $\mat{F}_\alpha=\mat{F}_\Gweiss$, i.e. the distribution 
is also configuration-independent.

The spectral functions for fixed $f$-particle configuration consist of subbands,
that can be identified with certain $K$-patches, as depicted in the third 
column of Fig.~\ref{fig:teff_conf}.
We observe that the plateaus in $h(\omega)$, between energy regions with different chemical potential,
occur at the boundaries between these subbands.
In particular, we can associate them with the region between the $(\pi,\pi)$ subband and the subbands corresponding to $K=(0,\pi),(\pi,0)$ in the configurations with one $f$-particle (second row),
and the region between the $(0,0)$ subband and the $K=(0,\pi),(\pi,0)$ subbands in the configurations with three $f$-particles (fifth row).

Finally, we plot the (artificially broadened) spectral function of the isolated $2{\times}2$ cluster at $\beta=2$ in the fourth column of Fig.~\ref{fig:teff_conf}.
The spectral peaks of the isolated plaquette can be clearly associated with the $K$-resolved spectral features of the embedded plaquette, although the latter are of course broadened due to lattice effects.

\section{Conclusions and outlook}
\label{sec:conclusions}

We have simulated interaction ramps in the 2D Falicov-Kimball model using a nonequilibrium implementation of DCA and compared the result for different clusters with up to $8$ sites. While these clusters are still too small to demonstrate a proper convergence of local and nonlocal expectation values with cluster size, we have shown that by averaging over different patch-layouts one can at least observe a systematic trend with cluster size (increasing correlations with increasing cluster size).

After a ramp to stronger interactions, the $c$ particles move away from the $f$ sites, which leads to strong nearest-neighbor $c$-$f$ correlations in the nonthermal steady state. These enhanced correlations are however a manifestation of the trapping in a nonthermal state, since the thermal system with the same energy would
have a more even distribution of $f$-particles and correspondingly weaker correlations. 
The opposite is true for the double occupancy, where the thermal state would exhibit a stronger reduction than the nonthermal steady state. By mapping out the differences between trapped and thermal expectation values for a range of initial and final interactions,
we found that 
for ramps within the metallic or insulating regime, the deviations from thermal behavior are relatively small, while ramps
across $U_c$ can lead to large deviations between the nonthermal steady-state value, and the thermal reference. 

Even in cases where the local or nearest-neigbor observables reach almost thermal values after the quench, the nonequilibrium energy distribution function can show large deviations from a thermal one. We defined an energy-dependent effective temperature from the ratio of the retarded and lesser Green's function and showed that even for quenches within the metal regime, there are considerable variations, especially near the edges of the subbands of the spectral function (associated with different $f$ particle occupations). Not even within the subbands it is possible to define a meaningful effective temperature, so that a description of the nonequilibrium steady state in terms of a few parameters ($f$-particle occupations, effective temperatures and effective chemical potentials) seems difficult.   

In the future, it would be interesting to extend this study to larger clusters using a Monte Carlo sampling of the initial $f$-particle configuration. Since the storage requirement of the nonequilibrium Green's functions is large, an explicit summation over all configurations, as done in this work, is not possible for substantially larger clusters. With clusters of size $8{\times}8$ or larger it would be possible to explore issues related to Anderson localization, since an interaction ramp from $U_0=0$ is equivalent to the switch-on of a disorder potential, and the Falicov-Kimball model has been shown to exhibit a rich phase diagram with an Anderson insulating phase near the Mott transition.\cite{antipov2016}

 
\begin{acknowledgments}
We thank L. Boehnke, D. Gole\v{z}, and H. Strand for helpful discussions.
The calculations were performed on the Beo04 cluster at the University of Fribourg.
AH and PW acknowledge support from ERC starting grant No. 278023. 
\end{acknowledgments}

\appendix
\section{Self-energy calculation}
\label{sec:solvesigma}
The impurity self-energy fulfills the Dyson equation in the following form:
\begin{equation}
  \label{eq:dyson}
  \mat{G} = \matGweiss + \mat{G}\mat{\Sigma}\matGweiss. 
\end{equation}
Additionally, we define a new contour function $\mat{X}$ which fulfills the 
following similar equation, and is also diagonal in $\Kcl$:
\begin{equation}
  \label{eq:dysonx}
  \mat{G} = \matGweiss + \matGweiss\mat{X}\matGweiss.
\end{equation}  
Comparison to \cref{eq:dyson} yields
\begin{equation}
  \mat{\Sigma} = \mat{G}^{-1}\matGweiss\mat{X},
\end{equation}  
whereas rearrangement yields
\begin{equation}
  \Id + \mat{X}\matGweiss = \matGweiss^{-1}\mat{G}.
\end{equation}  
The combination of the last two equations produces a contour Fredholm
equation of the second kind for the self-energy:
\begin{equation}
  \label{eq:selfen2}
  ( \Id + \mat{X}\matGweiss )\mat{\Sigma} = \mat{X}. 
\end{equation}

We still need to derive explicit forms for $\mat{X}$ and $\mat{X}\matGweiss$. 
To that end we insert the cluster solution Eqs.~(\ref{eq:cluster}),(\ref{eq:partial0}) into 
the Dyson equation (\ref{eq:dyson}), which yields
\begin{equation}
  \mat{G} = \matGweiss +
  \matGweiss (\Id + \sum_\alpha w_\alpha \mat{U}_\alpha \mat{R}_\alpha)
    \mat{\Sigma}\matGweiss. \
\end{equation}    
By comparison to Eqs.~(\ref{eq:dysonx}),(\ref{eq:selfen2}) we find
\begin{equation}  
  \mat{X}\matGweiss = \sum_\alpha w_\alpha \mat{U}_\alpha \mat{R}_\alpha,
\end{equation}  
and by applying $\matGweiss^{-1}$ from the right and using Eq.~(\ref{eq:partial0}) 
we find
\begin{equation}   
  \mat{X} = \sum_\alpha w_\alpha \mat{U}_\alpha \mat{R}_\alpha \mat{U}_\alpha.
\end{equation}

\section{Generalized Gibbs ensemble for Falicov-Kimball model}
\label{sec:gge_appendix}

Let us assume a Hamiltonian which can be written as a sum of conserved quantities $\hat I_\alpha$,
that commute with each other, 
\begin{align}
H&=
\sum_\alpha \epsilon_\alpha \hat I_\alpha,\quad [\hat I_\alpha,\hat I_\beta]=0.
\label{decompose}
\end{align}
Consequently, all the $\hat I_\alpha$ commute with $\hat H$,
\begin{align}
[H,\hat I_\alpha]
&=0.
\end{align}
In this situation, the generalized Gibbs ensemble (GGE) is given by the density matrix
\begin{align}
\rho_{\mathrm{GGE}}
&=
\frac{1}{Z_{\mathrm{GGE}}} e^{-\sum_\alpha \lambda_\alpha \hat I_\alpha},
\end{align}
where
\begin{align}
Z_{\mathrm{GGE}}
&=
{\mathrm{Tr}} (e^{-\sum_\alpha \lambda_\alpha \hat I_\alpha})
\end{align}
is the partition function for GGE, and $\lambda_\alpha$ are Lagrange multipliers.
If the system approaches the GGE in the long-time limit, then the $\lambda_\alpha$ are determined by the following set of conditions
\begin{align}
\langle \hat I_\alpha \rangle_{\mathrm{GGE}}
&=
\langle \hat I_\alpha \rangle_{t=+0},
\label{I=I}
\end{align}
since each $\hat I_\alpha$ is conserved during the time evolution.

\subsection{Falicov-Kimball model}

In the case of the Falicov-Kimball model, the Hamiltonian is given by
\begin{align}
H&=
-t\sum_{\langle i,j\rangle} (c_i^\dagger c_j+{\mathrm{h.c.}})
+U\sum_i \hat n_i^f \hat n_i^c-\mu_c\sum_i \hat n_i^c.
\end{align}
Here $\hat n_i^f$ is conserved for each $i$,
\begin{align}
[H,\hat n_i^f]
&=0.
\end{align}
This allows one to simultaneously diagonalize $H$ and $\hat n_i^f$.
In this basis, we can block-diagonalize the Hamiltonian in the form of (\ref{decompose}) as
\begin{align}
H&=
\sum_{\bm n_f}
\hat I_{\bm n_f}-\mu_c\sum_{\bm n_f} \hat N_{\bm n_f},
\\
\hat I_{\bm n_f}
&=
\mathcal P_{\bm n_f}
\left(
-t\sum_{\langle i,j\rangle} (c_i^\dagger c_j+{\mathrm{h.c.}})
+U\sum_i n_i^f \hat n_i^c
\right),
\\
\hat N_{\bm n_f}
&=
\mathcal P_{\bm n_f}
\sum_i \hat n_i^c,
\end{align}
where
\begin{align}
\sum_{\bm n_f}
&=
\sum_{n_1^f=0,1} \sum_{n_2^f=0,1}
\cdots
\sum_{n_N^f=0,1}
\end{align}
($N$ is the number of lattice sites), and
\begin{align}
\mathcal P_{\bm n_f}
&=
|n_1^f, n_2^f,\dots\rangle \langle n_1^f, n_2^f, \dots|
\end{align}
is a projection operator onto the eigenspace of $\bm n_f=\{\hat n_i^f\}$.
It is easy to see that
\begin{align}
[\hat I_{\bm n_f},\hat I_{\bm n_f'}]
&=0,
\\
[\hat I_{\bm n_f},\hat N_{\bm n_f'}]
&=0,
\\
[\hat N_{\bm n_f},\hat N_{\bm n_f'}]
&=0,
\end{align}
since
\begin{align}
\mathcal P_{\bm n_f}
\mathcal P_{\bm n_f'}
&=0
\quad
(\bm n_f \neq \bm n_f').
\end{align}

\subsection{Interaction quench}

Let us consider a situation where the interaction parameter $U$ is quenched as $U=U_- \to U_+$
at $t=0$. Correspondingly, we define
\begin{align}
\hat I_{\bm n_f}^\pm
&=
\mathcal P_{\bm n_f}
\left(
-t\sum_{\langle i,j\rangle} (c_i^\dagger c_j+{\mathrm{h.c.}})
+U_\pm\sum_i n_i^f \hat n_i^c
\right).
\end{align}
Since $\hat I_{\bm n_f}^\pm$ and $\hat N_{\bm n_f}$ are quadratic
in the fermionic operators and commute with each other,
we can further diagonalize them with single-particle eigenstates $|\bm n_f,\alpha_\pm\rangle$
and eigenvalues $\varepsilon_{\bm n_f, \alpha_\pm}$, where $\alpha_\pm$ labels each eigenstate
before and after the quench. In this basis, we can write
\begin{align}
\hat I_{\bm n_f}^\pm
&=
\mathcal P_{\bm n_f} \sum_{\alpha_\pm} \varepsilon_{\bm n_f,\alpha_\pm} \hat n_{\alpha_\pm}^c,
\\
\hat N_{\bm n_f}
&=
\mathcal P_{\bm n_f} \sum_{\alpha_\pm} \hat n_{\alpha_\pm}^c,
\end{align}
where $\hat n_{\alpha_\pm}^c=c_{\alpha_\pm}^\dagger c_{\alpha_\pm}$.
We can see that
\begin{align}
\hat N_{\bm n_f,\alpha_+}
&=
\mathcal{P}_{\bm n_f} \hat n_{\alpha_+}^c
\end{align}
is conserved for each $\bm n_f, \alpha_+$ after the quench.
$\hat N_{\bm n_f,\alpha_+}$ are the finest conserved quantities.
$\hat I_{\bm n_f}^+$ is linearly dependent on them
($\hat I_{\bm n_f}^+=\sum_{\alpha_+} \varepsilon_{\bm n_f,\alpha_+} \hat N_{\bm n_f,\alpha_+}$).

The most general GGE is
\begin{align}
\rho_{\mathrm{GGE}}
&=
\frac{1}{Z_{\mathrm{GGE}}}e^{-\sum_{\bm n_f,\alpha_+} \lambda_{\bm n_f,\alpha_+}\hat N_{\bm n_f,\alpha_+}},
\end{align}
where $\lambda_{\bm n_f,\alpha_+}$ is the Lagrange multiplier. The constraint on it is given by
Eq.~(\ref{I=I}), which reads in the present case
\begin{align}
\langle \hat N_{\bm n_f,\alpha_+}\rangle_{\mathrm{GGE}}
&=
\langle \hat N_{\bm n_f,\alpha_+}\rangle_{t=+0}.
\end{align}
One can calculate both sides explicitly as
\begin{align}
\frac{1}{e^{\lambda_{\bm n_f,\alpha_+}}+1}
&=
\sum_{\alpha_-} |\langle \bm n_f, \alpha_+|\bm n_f, \alpha_-\rangle|^2 f(\varepsilon_{\bm n_f,\alpha_-}),
\label{lambda}
\end{align}
where $f(\varepsilon)=1/(e^{\beta(\epsilon-\mu)}+1)$ is the initial thermal fermi distribution.
This completely determines $\lambda_{\bm n_f,\alpha_+}$.

\nocite{tange2011}
\bibliography{literature}

%
\end{document}